\documentclass[namedreferences,hyperref,optionalrh]{article}
\usepackage[T1]{fontenc}
\usepackage{graphicx}  
\usepackage{color} 
\usepackage{xcolor}
\usepackage{soul}
\usepackage{amsmath}
\usepackage{natbib}
\usepackage{newtxtext}
\usepackage{tabto}
\usepackage[affil-it]{authblk}

\chardef\us=`\_

\title{Using the Schmidt Decomposition to Determine Quantum Entanglement}
\author{Lane Boswell and Ying Cao}
\affil{Drury University}
\begin{document}

\maketitle
\begin{abstract}
Quantum information theory is a rapidly growing area of math and physics that combines two independent theories, quantum mechanics and information theory. Quantum entanglement is a concept that was first proposed in the EPR paradox. In quantum mechanics, particles can be in superposition, meaning they are in multiple different states at once. It is not until the particle is measured that it is forced into a single state. However, it is possible that particles can be tied to other particles, meaning that the measurement of one particle will determine the measurement of the other particle. Entanglement is at the very core of quantum information theory. It is one of the core pieces that allows for the massive increase in computing power. For this paper, we decided to focus on demonstrating the mathematical method (the Schmidt Decomposition) for determining if a system is entangled, and a demonstration of quantum entanglement’s use (quantum teleportation) as well as a quick look at how to extend the uses of the Schmidt decomposition.  
\end{abstract}
Keywords: Qubit, Schmidt decomposition; Entanglement; Singular value decomposition; Partial trace\\
Email address: lboswell@drury.edu

\section{Introduction to Quantum Information}
     \label{S-Introduction} 

\tab Quantum information theory is a rapidly growing area of math and physics that combines two independent theories, quantum mechanics and information theory. The field applies ideas found within quantum mechanics to classical information theory, creating a more capable and efficient system. Quantum information theory poses many advantages to classical information theory, which mainly revolve around their efficiency and fundamentally novel cryptography. The novelty comes from quantum mechanics concepts such as superposition, interference, and entanglement. \\
\indent This paper will explore how quantum entanglement can be determined using the Schmidt decomposition, along with extensions and applications of this mathematical tool. The examples used in this paper will primarily be about bipartite systems (i.e., systems of two qubits).

\section{{The Schmidt decomposition}} 
      \label{S-general}  
\subsection{Conceptual Idea}
\hspace{5mm}Schrodinger's definition of entanglement states that an entangled pure state is a pure quantum state of multipartite systems that cannot be represented in the form of a simple tensor product of subsystem states (\cite{jaeger2007quantum}). A demonstration of a simple tensor product is shown below. 
\begin{center}
    \(|\psi>=|0> \otimes |1>\)
\end{center}

Therefore, looking at a pure bipartite state, if the system is decomposed into its simplest form it will either have one tensor product, or it will be the sum of multiple tensor products. Using the Schmidt decomposition we can determine whether a system is entangled or not so long as it is a pure bipartite system. The Schmidt decomposition is a way of analyzing a Hilbert space that is a tensor product of two Hilbert subspaces.
\begin{center}
    \(H^{AB}=H^A \otimes H^B\)
\end{center}
The two subsystems are commonly referred to as Alice and Bob, or simply A and B. When we perform the Schmidt Decomposition we are performing a change of basis on our quantum state to see if we can represent the state as a single tensor product between a pure state of subsystem A and a pure state of subsystem B. If this is possible we say that the quantum state is separable. However, if our quantum state must be written as a sum of tensor products between pure states of subsystem A and subsystem B, then the quantum system is said to be entangled. In practice this is done by finding the Schmidt coefficients through the Schmidt decomposition. we define the Schmidt number to be the total number of nonzero Schmidt coefficients. If the Schmidt number is 1 then the multi-qubit system is separable. If the system has a Schmidt number greater than 1, then the system is entangled.

\subsection{First Method: Singular Value Decomposition}

\indent For the first method start by writing the multiple quantum bit (multi-qubit) system out as a sum of tensor products with attached coefficients.
\begin{center}
    \(|Z>=\sum\limits_{i=0}^{N-1}\sum\limits_{j=0}^{M-1}C_{ij}|\psi_i> \otimes |\phi_j>\)
\end{center}
Where N and M are the size of the dimensions of \(H^A\) and \(H^B\) respectively, and \(|\psi_i>\) and \(|\phi_j>\) are the sets of basis vectors for \(H^A\) and \(H^B\) respectively. Note that any basis can be chosen for this first step. \\

From here construct the matrix of correlations as follows:
\begin{center}
    \(C_{AB}=
    \begin{bmatrix}
        C_{0,0} & C_{0,1} & ... & C_{0,(M-1)} \\
        C_{1,0} & ... & ... & C_{1,(M-1)} \\
        ... & ... & ... & ... \\
        C_{(N-1),1} & ... & ... & C_{(N-1),(M-1)}
    \end{bmatrix}\)
\end{center}
We will now perform singular value decomposition on this matrix to determine entanglement. \\

singular value decomposition is a way of decomposing a matrix into three separate matrices of the form:
\begin{center}
    \(C_{AB}=U\Sigma V^T\)
\end{center}
Where U is the set of left singular vectors, \(V^T\) is the set of right singular vectors, and \(\Sigma\) is a diagonalized matrix with the singular values along the diagonal in decreasing order. In the context of our multi-qubit system, U is the set of basis vectors for subsystem A, and V is the set of basis vectors for subsystem B. The singular values are the coefficients that go in front of the tensor product between its corresponding left and right singular vectors. These singular values turn out to be the Schmidt coefficients. The coefficients relate to the probability of measuring the multi-qubit system in each ones corresponding state. \\

Now we can rewrite the multi-qubit system as a single summation instead of a double summation. \(|u_k>\) and \(|v_k>\) is the the columns of U and V respectively (Note here we use the matrix V and not \(V^T\)), \(\sigma_k\) is the set of singular values in decreasing order, and S is the total number of nonzero singular values (the Schmidt number)
\begin{center}
    \(|Z>=\sum\limits_{k=1}^{S}\sigma_k|u_k>\otimes |v_k>\)
\end{center}
This yields the completed Schmidt decomposition. The resulting form is the simplest we can represent a pure bipartite system in
\subsection{Example Using The Singular Value Decomposition Method}
\ \ \ \ Start with a two-qubit system in the following state:
\begin{center}
    \(|Z>=\frac{1}{2}(|00>+|01>+|10>+|11>)\)
\end{center}
This system is in a superposition of all four possible states in the computational basis with even probability in each. separating each state into it's subsystems we have:
\begin{center}
    \(|Z>=\frac{1}{2}(|0>_A|0>_B+|0>_A|1>_B+|1>_A|0>_B+|1>_A|1>_B)\)
\end{center}
We have not made any manipulations to the system, just simply indicated the subsystem that each element of a state belongs to. Now we can easily construct a matrix of correlations:
\begin{center}
    \(C_{AB}=\frac{1}{2}
    \begin{bmatrix}
        1 & 1 \\
        1 & 1
    \end{bmatrix}\)
\end{center}
From here we can perform singular value decomposition to find our Schmidt coefficients as well as our basis vectors for subsystems A and B
\begin{center}
    \(\sigma_1=1, \ \sigma_2=0\) \\
    \(v_1= \frac{1}{\sqrt{2}}\begin{bmatrix}
        1 \\
        1
    \end{bmatrix}\) \\
    \(v_2=\frac{1}{\sqrt{2}}\begin{bmatrix}
        1 \\
        -1
    \end{bmatrix}\) \\
    \(u_1=\frac{1}{\sqrt{2}}\begin{bmatrix}
        1 \\
        1
    \end{bmatrix}\) \\
    \(u_2=\frac{1}{\sqrt{2}}\begin{bmatrix}
        1 \\
        -1
    \end{bmatrix}\) 
\end{center}
Using the formula \(C_{AB}=U\Sigma V^T\) it can be seen that this is indeed the correct decomposition of our matrix of correlations. 
 \\

 From the decomposition we can determined that the Schmidt number is one, as there is only one nonzero Schmidt coefficient. Thus, this two-qubit system is separable. The qubit can now be rewritten in the simplified form:
 \begin{center}
     \(|Z>=1\begin{bmatrix}
        \frac{1}{\sqrt{2}}\\
        \frac{1}{\sqrt{2}}
    \end{bmatrix} \otimes
    \begin{bmatrix}
        \frac{1}{\sqrt{2}} \\
        \frac{1}{\sqrt{2}}
    \end{bmatrix}\)
 \end{center}
However, the vector \(\begin{bmatrix}
        \frac{1}{\sqrt{2}} \\
        \frac{1}{\sqrt{2}}
    \end{bmatrix}\) 
is also a basis vector in the diagonal basis given the name \(|+>\). So we can more succinctly say
\begin{center}
    \(|Z>=|++>\)
\end{center}

\subsection{Second Method: Partial Trace}
\ \ \ \ For the second method, we start by constructing the density matrix for a multi-qubit system. This is done by taking the outer product of the multi-qubit system with itself:
\begin{center}
    \(\rho_{AB}=|Z><Z|\)
\end{center}

From this density matrix of our bipartite system we can now find the reduced density matrices of both \(H^A\) and \(H^B\) by performing partial traces. The formulas
\begin{center}
    \(\rho_A=tr_B[\rho_{AB}]\) \\
    \(\rho_B=tr_A[\rho_{AB}]\)
\end{center}
will give the two reduced density matrices. To preform the partial trace in Bra-ket notation we simply ignore the subsystem we are not taking a trace of and perform a regular trace on the other subsystem:
\begin{center}
    \(tr_A[(|\psi_1>_A|\phi_1>_B)(<\psi_1|_A<\phi_2|_B)]=|\phi_1><\phi_2|(tr[|\psi_1><\psi_2|])\)
\end{center}

Next we find the eigenvalues and eigenvectors of both reduced density matrices. The eigenvalues of both matrices will be the same (if one matrix is higher dimensional, then it will have all the eigenvalues of the other, and the rest will be 0) but the eigenvector may differ. Square rooting the eigenvalues will give us the Schmidt coefficients of our original multi-qubit system (denoted \(\sigma_k\)). The eigenvectors of \(\rho_A\) and \(\rho_B\) will be the basis vectors for \(H^A\) and \(H^B\) respectively (denoted \(|u_k>\) and \(|v_k>\) respectively). Finally, we reconstruct the system the same way we did before in method 1, to give us the Schmidt decomposition:
\begin{center}
    \(|Z>=\sum\limits_{k=1}^{S}\sigma_k|u_k>\otimes |v_k>\)
\end{center}
where once again, S is the number of nonzero \(\sigma_k\)s
\subsection{Example Using The Partial Trace Method}
\tab This time we will look at a three-qubit system. This will be helpful for the upcoming section, quantum teleportation, as it's applications arise from a three-qubit system.
\begin{center}
    \(|w>=\frac{1}{\sqrt{3}}(|001>+|010>+|100>)\)
\end{center}
Start by composing the multi-qubit system's density matrix, where we will partition it into a 1-qubit subsystem tensored with a two-qubit subsystem:\\
\begin{center}
    
\(|w> \ =\frac{1}{\sqrt{3}}(|0>_A|01>_B+|0>_A|10>_B+|1>_A|00>_B)\) \\\end{center}
\(\rho_{AB} \ =|w><w|\) \\
\indent \indent \(=\frac{1}{3}(|0>_A|01>_B+|0>_A|10>_B+|1>_A|00>_B)\)\\ \indent \indent\(\ \ \ \ \ \ \ (<0|_A<01|_B+<0|_A<10|_B+<1|_A<00|_B)\)\\ \\
\(\rho_{AB} =\frac{1}{3}([|0><0|]_A[|01><01|]_B+[|0><0|]_A[|01><10|]_B+\)\\ \\
\indent \indent \(\ \ \ \ \ \ \ [|0><1|]_A[|01><00|]_B+[|0><0|]_A[|10><01|]_B+\)\\ \\
\indent \indent \(\ \ \ \ \ \ \ [|0><0|]_A[|10><10|]_B+[|0><1|]_A[|10><00|]_B+\)\\ \\
\indent \indent \(\ \ \ \ \ \ \ [|1><0|][|00><01|]_B+[|1><0|]_A[|00><10|]_B+\)\\ \\
\indent \indent\(\ \ \ \ \ \ \ [|1><1|]_A[|00><00|]_B)\) \\

This is quite an unruly set of terms. However, when we go to find the two reduced density matrices we can use two helpful traits. One is that the trace of an outer product is equal to the corresponding inner product:
\begin{center}
    \(tr[|\psi><\phi|]=<\phi|\psi>\)
\end{center}
Secondly, both subsystems are constructed with an orthonormal basis, therefore all inner products will be of the form:
\begin{center}
    \(<\phi|\psi>=\delta_{\psi,\phi}\); where \
    \(\begin{cases}
        \delta_{\psi,\phi}=1 & \text{if} \ \psi=\phi \\
        \delta_{\psi,\phi}=0 & \text{if} \ \psi \neq \phi
    \end{cases}\)
\end{center}
This means when we take the partial trace, we only need to look at the terms where we're taking the trace of a state with itself, since all other terms become 0, and the trace of these states with themselves will all equal 1. This leads to much simpler representations of the reduced density matrices:
\begin{center}
    \(\rho_A=\frac{1}{3}(|0><0|+|0><0|+|1><1|)\) \\
    \(\rho_B=\frac{1}{3}(|01><01|+|01><10|+|10><01|+|10><10|+|00><00|)\)
\end{center} 
Writing this out in matrix form gives us:
\begin{center}
    \(\rho_A=
    \begin{bmatrix}
        \frac{2}{3} & 0 \\
        0 & \frac{1}{3}
    \end{bmatrix}\), \ 
    \(\rho_B=
    \begin{bmatrix}
        \frac{1}{3} & 0 & 0 & 0 \\
        0 & \frac{1}{3} & \frac{1}{3} & 0 \\
        0 & \frac{1}{3} & \frac{1}{3} & 0 \\
        0 & 0 & 0 & 0
    \end{bmatrix}\)
\end{center}

Now all that's left to do is find eigenvalues and eigenvectors. It is clear to see that the eigenvalues for \(\rho_A\) are 2/3 and 1/3. This means \(\rho_B\)'s must be 2/3, 1/3, and 0 with a multiplicity of 2. Therefore the Schmidt coefficients and their corresponding eigenvectors are: \\
\begin{center}
    
\(\sigma_1=\sqrt{\frac{2}{3}}, \ \frac{1}{\sqrt{3}}\) \\
\(|u_1>=
\begin{bmatrix}
    1 \\
    0
\end{bmatrix},\ 
|u_2>=\begin{bmatrix}
    0 \\
    1
\end{bmatrix}\) \\
\(|v_1>=\begin{bmatrix}
    0 \\
    \frac{1}{\sqrt{2}} \\
    \frac{1}{\sqrt{2}} \\
    0
\end{bmatrix}, \ 
|v_2>=\begin{bmatrix}
    1 \\
    0 \\
    0 \\
    0
\end{bmatrix}\)\end{center}
\newpage
Now that we have all the needed components, we can reconstruct our system into the following form:
\begin{center}
\(|w>=\sqrt{\frac{2}{3}}\begin{bmatrix}
    1 \\
    0
\end{bmatrix} \otimes
\begin{bmatrix}
    0 \\
    \frac{1}{\sqrt{2}} \\
    \frac{1}{\sqrt{2}} \\
    0
\end{bmatrix} +
\frac{1}{\sqrt{3}}
\begin{bmatrix}
    0 \\
    1
\end{bmatrix} \otimes
\begin{bmatrix}
    1 \\
    0 \\
    0 \\
    0
\end{bmatrix}\)
\end{center}
Since the simplest form of our bipartite system requires us to sum together two separate tensor products we say that our three-qubit system is entangled. \\
\indent From the past few sections it is clear to see the Schmidt decomposition indicates an interesting mathematical connection between singular value decomposition and partial traces. This connection is further explored by Tai-Danae Bradley in her paper \textit{At the interface of Algebra and Statistics} 
(\cite{bradley2020interface}).
\section{Application: Quantum Teleportation}
\tab In Quantum teleportation (\cite{jaeger2007quantum}) the desired goal is to transmit the 
information stored in a qubit from one location, which we'll call lab 
A, to another location called lab B, without physically transporting 
the qubit. This will be done by having a qubit in lab B that we will 
reconstruct into the exact same state as the original qubit. To do this 
we start by having a two qubit system that is entangled:
\(|\phi^+>=\frac{1}{\sqrt{2}}|00>+|11>\) \\
One of these qubits is held in lab A, while the other is held in lab B. 
Then we have another qubit in lab A in some unknown state:\\
\begin{center}
   \(|\psi>=a_0|0> +a_1|1>\)  
\end{center}
This is the qubit that we wish to reconstruct in lab B. To do this we 
can formulate a three-qubit system:\\
\begin{center}
    
\(|\Psi>=|\psi>|\phi^+>=\frac{1}{\sqrt{2}}(a_0|0>+a_1|1>)(|00>+|11>)\)
\end{center}
This can be rewritten as: \\
\(|\Psi>=\frac{1}{2}(|00>+|11>)(a_0|0>+a_1|1>) \\ \\
\indent \ \ \ +\frac{1}{2}(|00>-|11>)(a_0|0>-a_1|1>)\\ \\
\indent \ \ \ +\frac{1}{2}(|01>+|10>)(a_0|1>+a_1|0>)\\ \\
\indent \ \ \ +\frac{1}{2}(|01>-|10>)(a_0|1>-a_1|1>)\) \\

Where the two-qubit system on the left is both of lab A's qubit's and 
the 
1-qubit system on the right is the lab B qubit. No physical action was 
taken here this is simply an algebraic rearrangement of our system. The 
rearrangement allows us to see that the three-qubit system is in a 
superposition between 4 states (These 4 states make up the Bell basis, 
a basis for entangled two-qubit systems). If we perform what is known 
as 
a Bell measurement on the two qubits in lab A then we will know which 
of the 4 states lab B's qubit is in. Now all we have to do is 
communicate this information (which is 2 classical bits worth of 
information) to lab B so that they know what state their qubit is in. 
From here they can perfectly reconstruct the original qubit's
(\(|\psi>\)) form using one of the 4 Pauli matrices:\\
\begin{center}
\(I=
\begin{bmatrix}
    1 & 0 \\
    0 & 1
\end{bmatrix}, \
X=
\begin{bmatrix}
    0 & 1 \\
    1 & 0
\end{bmatrix}, \ 
Y=\begin{bmatrix}
    0 & -i \\
    i & 0
\end{bmatrix}, \ 
Z= 
\begin{bmatrix}
    1 & 0 \\
    0 & -1
\end{bmatrix}\)
\end{center}

\section{Extensions of the Schmidt Decomposition}
\tab Another approach to entanglement is through constructing an entanglement \\witness. According to Horodecki, a density matrix \(\rho\) on \(H^A \otimes H^B\) is entangled if and only if there exists a Hermitian matrix W, called an entanglement witness, such that its expectation value is positive for every separable state but negative for some entangled state (\cite{horodecki1997separability}). Specifically,  \\
\begin{center}
    \(tr[W\rho] <0\) \\
    \(tr[W\rho_i]>0\) \end{center}
for all separable states \(\rho_i\). The problem now is being able to construct such a matrix. To perform this, we can use a version of the Schmidt decomposition called the operator Schmidt decomposition to construct such a matrix.\\
\indent The Schmidt decomposition can be done in the operator space of an operator X the same way as it does on a pure bipartite state. We once again find the Schmidt coefficients of X and list them in decreasing order. We can now reconstruct the operator X as follows: \\
\begin{center}
    \(X=\sum\limits_i^S\mu_iG^A_i \otimes G^B_i\)
\end{center}
The entanglement witness W, then can be constructed from operator X as follows: \\
\begin{center}
    \(W=\mu_1I-X\) \end{center}
Where \(\mu_1\) is the largest Schmidt coefficient of the operator Schmidt decomposition of X, and I is the identity matrix (\cite{zhang2024analyzing}). \\
\indent The expectation value of the entanglement witness operator W can be determined based on the Schmidt decomposition on X (\cite{zhang2024analyzing}). When operator X is the density matrix, the operator Schmidt decomposition reduces to the traditional Schmidt decomposition on quantum states. For general operators (not just states), operator Schmidt decomposition gives a broader class of witnesses. \\
\indent The entanglement witness approach can address multiple-particle systems and can evaluate different degree of entanglement. It can solve entangled mixed state and has application in quantum computing. All these areas extend the scope of discussion about entanglement of quantum states. 

\section{Conclusion}
\tab The Schmidt decomposition is a powerful tool in quantum 
information theory that allows us to gain a better understanding of a 
system. By applying the Schmidt decomposition we are able to both 
determine whether a bipartite system is entangled, and represent the 
system in the simplest form. We can utilize this knowledge in quantum 
teleportation by determining whether a two-qubit system is entangled. 
Once we have guaranteed entanglement between the two qubits, we can now 
perform teleportation based on it. Without this tool we would be 
unable to determine if two qubits were entangled, leaving the foundation to operationalize quantum teleportation unknown. \\
\indent The discussion in the paper focuses on bipartite systems. However, Schmidt decomposition does not necessarily have this limitation. As seen in the 
extensions we can also go further with the Schmidt decomposition to 
develop a general tool that allows different approaches to determining 
quantum entanglement in multi-particle systems. 
\newpage

\bibliographystyle{plainnat}
\bibliography{SD}
\end{document}